\documentclass[superscriptaddress,floatfix,prl,twocolumn,amsmath,amssymb]{revtex4-2}

\usepackage{graphicx}
\usepackage{dcolumn}
\usepackage{bm}
\usepackage{amssymb}
\usepackage{amsfonts}
\usepackage{natbib}
\usepackage{amsmath}
\usepackage{mathtools}
\usepackage{xcolor}
\usepackage{color}
\usepackage{datetime}
\usepackage{footnote}
\usepackage{bbold}
\usepackage[normalem]{ulem}
 \usepackage{ragged2e}
 \usepackage{xfrac}
 \usepackage{sidecap}
 \usepackage{setspace}
 \usepackage{multirow}
 \usepackage{xcolor}
 \usepackage{flushend}
\usepackage{hyperref}
\usepackage{cleveref}
\usepackage{microtype}
\usepackage{siunitx}

\usepackage{braket}
\usepackage{nicefrac, xfrac}

\newcommand{\orange}[1] {\textcolor{orange}{#1}}
\newcommand{\blue}[1] {\textcolor{blue}{#1}}

\usepackage{stackengine,xcolor}
\def\cpm{\mathbin{\ensurestackMath{\abovebaseline[-3.4pt]{%
  \stackunder[-3.5pt]{\color{blue!70}+}{\color{orange}-}}}}}
\def\cmp{\mathbin{\ensurestackMath{\abovebaseline[-1.5pt]{%
  \stackunder[-3.6pt]{\color{blue!70}-}{\color{orange}+}}}}}

\begin{document}

\author{Tim Weiss}
\address{Quantum Photonics Laboratory and Centre for Quantum Computation and Communication Technology, RMIT University, Melbourne, VIC 3000, Australia}

\author{Akram Youssry}
\address{Quantum Photonics Laboratory and Centre for Quantum Computation and Communication Technology, RMIT University, Melbourne, VIC 3000, Australia}

\author{Alberto Peruzzo}
\address{Quantum Photonics Laboratory and Centre for Quantum Computation and Communication Technology, RMIT University, Melbourne, VIC 3000, Australia}
\address{Quandela, Massy, France}

\title{Quantum nonlinear parametric interaction in realistic waveguides: a comprehensive study}

\begin{abstract}
Nonlinear sources of quantum light are foundational to nearly all optical quantum technologies and are actively advancing toward real-world deployment. Achieving this goal requires fabrication capabilities to be scaled to industrial standards, necessitating precise modeling tools that can both guide device design within realistic fabrication constraints and enable accurate post-fabrication characterization. In this paper, we introduce a modeling framework that explicitly integrates the engineering tools used for designing classical properties of integrated waveguides with quantum mechanical theory describing the underlying nonlinear interactions. We analyze the validity and limitations of approximations relevant to this framework and apply it to comprehensively study how typical fabrication errors and deviations from nominal design---common in practical waveguide manufacturing---affect the nonlinear optical response. Our findings highlight, in particular, a critical sensitivity of the framework to group-velocity dispersion, the potentially disruptive role of geometric inhomogeneities in the waveguide, and an upper bound on single-mode squeezed-state generation arising from asymmetric group-velocity matching conditions.


\end{abstract}

\maketitle

\bigskip

\section{I. Introduction}
Optical nonlinear interactions have played a pivotal role in advancing both cutting-edge technologies and fundamental research for over half a century. Today, one of their most promising frontiers lies in the generation of quantum light---a cornerstone for the development of optical quantum technologies. This application is now poised to transition from proof-of-principle demonstrations to deployable, application-grade systems \cite{ZhongPan:20,YinPan:20,JiaMalalvala:24,VanLeentWinfurter:22}. Achieving this transition will hinge on the ability to support high-quality, industrial-scale upscaling.

In the context of quantum optical nonlinear interactions, this upscaling is often envisioned as a shift from bulk-crystal implementations to integrated waveguide platforms. These offer enhanced scalability, improved stability, and, thanks to their reduced interaction volumes, significantly increased nonlinear interaction \cite{WangThompson:20,FiorentinoMunro:07}. Realizing application-grade waveguide technologies, however, demands accurate device modeling that captures both the classical characteristics of the waveguides and the quantum mechanical interactions occurring within them.

In this paper, we address this challenge from two complementary perspectives:

\noindent
(\textit{1}) We introduce a modeling framework that integrates classical engineering tools for waveguide design with theoretical methods capable of describing quantum nonlinear parametric interactions. We provide a comprehensive analysis of the approximations necessary to bridge these domains. Our focus is on quantum three-wave mixing processes mediated by the $\chi^{(2)}$ nonlinearity---processes that are both highly tunable through classical waveguide engineering and acutely sensitive to imperfections in the same.

\noindent
(\textit{2}) Building on this framework, we conduct a detailed study of pulsed quantum nonlinear interactions in realistic waveguides, explicitly accounting for fabrication-induced imperfections and deviations from ideal designs. We quantitatively analyze how these nonidealities---such as scattering losses at the waveguide surface, errors in domain inversions required for phase matching, and inhomogeneities introduced during waveguide and wafer fabrication \cite{LukeZhang:20,ChenFan:24}---affect the generation of quantum states. Specifically, we consider their impact on low- and high-gain generation of heralded pure states and single-mode squeezed states \cite{PickstonFedrizzi:21,HoudeQuesada:23}, broadband squeezed vacuum generation \cite{NehraMarandi:22,KashiwazakiFurusawa:20}, and quantum frequency conversion \cite{MaringRiedmatten:18,AnsariSilberhorn:18}. An overview of these effects and the corresponding nonlinear processes is provided in Fig. 1.

Through this investigation, we identify several key findings. Most notably, inhomogeneities in the waveguide geometry can have a highly disruptive impact on the generated quantum states. We also observe that,  at times, there's a critical dependence of the accuracy of the modeling framework on the explicit inclusion of group-velocity dispersion. Furthermore, we quantify an upper limit to the purity and single-modeness of squeezed-state generation from asymmetric group-velocity matching conditions, arising from self-phase modulation of the pump field.

The remainder of this paper is organized as follows: Section II provides a brief introduction to quantum optical nonlinear interaction in waveguides and outlines the key approximations used in our model. Section III presents the modeling framework in detail. Section IV delivers a quantitative analysis of the effects of fabrication errors and deviations from ideal designs. Section V explores the limiting case of pump-induced self- and cross-phase modulation. Finally, Section VI discusses the broader implications of our findings and outlines directions for future work.

\section{II. Waveguided nonlinear interaction}

\begin{figure*}
    \centering
    \includegraphics[width=2\columnwidth]{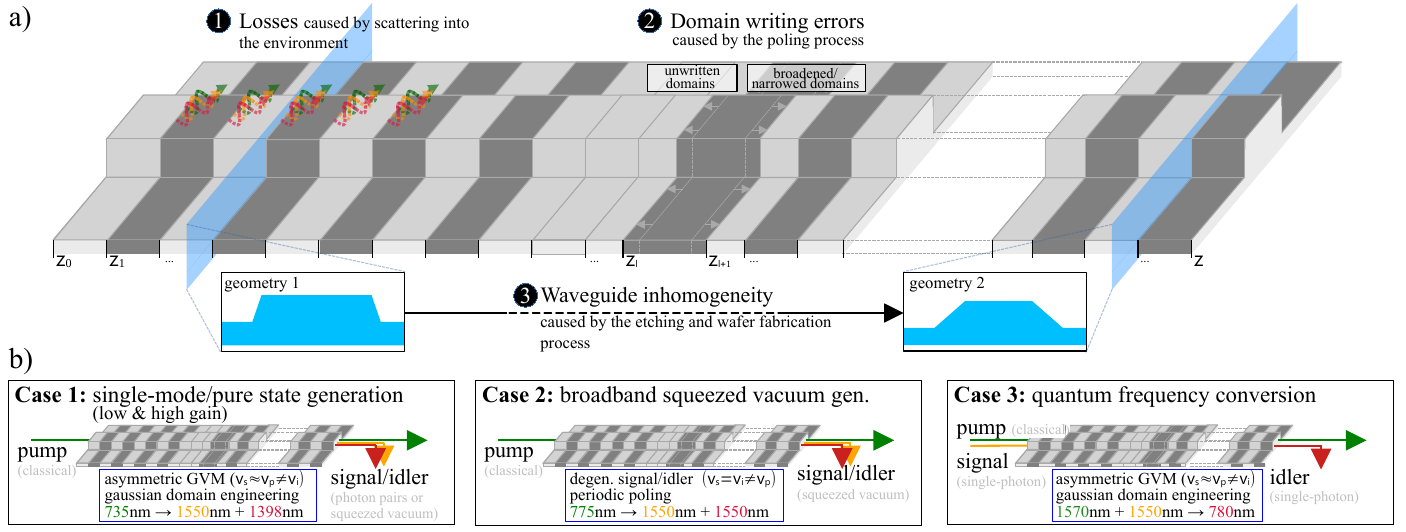}
    \caption{\textbf{a}) Schematic depiction of a ridge waveguide affected by different effects associated with realistic waveguides. Dark regions represent areas where the material nonlinearity was inverted artificially. The coordinates $z_{l}$ correspond to the positions of the Trotter-Suzuki expansion described in Sec. III. \textbf{b}) Schematic depiction of the different processes considered for the quantitative analysis in Sec. IV and V, highlighting in- and outputs, the interacting wavelengths, group-velocity matching conditions, and the type of domain patterning.}
    \label{fig:ConceptualFig}
\end{figure*}

Second order optical nonlinear interaction can, generally speaking, be separated into two different processes: a down-conversion process generating two distinct photons from a single photon of the input field - from hereon referred to as parametric down-conversion (PDC) \cite{Couteau:18,TriginerWalmsley:20} - and a frequency conversion process combining two input photons into a single output, referred to as quantum frequency conversion (QFC) \cite{RaymerSrinivasan:12}. While both processes are fundamentally quantum mechanical, they can, under certain conditions, be made to produce classical outputs. The three interacting fields are referred to as pump, signal and idler.

To implement a conversion process between fields of specific properties (e.g. in terms of wavelength or spatial modes), both conversation of energy and momentum need to be satisfied. The latter of these, also referred to as phase-matching, requires, in waveguides, artificial structuring of the material by creating locally constrained domains in which the nonlinearity in inverted - a process referred to as poling \cite{HumFejer:07}. If these requirements are satisfied, the conversion processes are characterized primarily by the dispersion properties of the interacting fields, given by both the intrinsic dispersion of the material and the confinement in the waveguide, the latter of which relates directly to its geometry. We note, that both the artificial structuring of the material nonlinearity and the dispersion imparted by the waveguide represent engineering angles which can be used to design the conversion processes \cite{WeissPeruzzo:25,JankowskiFejer:21}.

Accordingly, the interaction is ultimately defined by the properties of the interacting fields, the artificial structuring of the material enabling the interaction, and the dispersion at the respective wavelengths. In a realistic device, each of these quantities is affected by errors, rooted in the waveguide fabrication processes, and will deviate from the design along the length of the device. In this paper we separate the effects associated with such deviations from design into (1) the quality of confinement of light in the waveguide, corresponding to (linear) losses, (2) errors in the domain writing process, both in terms of an undesired domain broadening/narrowing, and domains that are not written at all, and (3) variations of the waveguide geometry and the associated waveguide-dispersion, rooted in the etching and wafer fabrication process (see also Fig. \ref{fig:ConceptualFig}a).

For the quantitative analysis in Sec. IV and V, we consider explicitly a Lithium Niobate ridge waveguide (see Table. \ref{tbl:Table}), which, arguably, represents one of the primary candidates for technological upscaling of nonlinear quantum optics, featuring ultra-low loss waveguides, mature waveguide fabrication, commercially available thin-film wafers, and strong nonlinear interaction including post-fabrication domain writing \cite{ZhuLoncar:21}. Lithium Niobate, however, does feature dispersion properties suited for symmetric group-velocity matching (GVM) in the visible/near-infrared region and relies instead on dispersion-engineered asymmetric GVM for the generation of heralded pure stats and single-mode squeezed-states \cite{URenRaymer:06}. As we show below, this ultimately limits the maximal amount of squeezing available to such states.

We will, to end this section, shortly summarize the approximation used in our modeling framework:

\textbf{Taking approximations:} Our model relies, first and foremost, on the parametric approximation, pertaining to a pump field undepleted by the nonlinear interaction, allowing it to be modeled classically \cite{QuesadaSipe:22,JankowskiFejer:24}. This is generally applicable if the number of pump photons is much larger than the number of photons in the signal and idler fields, and results in equations of motion linear in creation and annihilation operators, describing a gaussian evolution. We further assume the different nonlinear interaction coefficients to be constant within pump, signal and idler frequency ranges and take them to be unaffected by the waveguide inhomogeneity. We similarly take the group velocity to be position independent, but, critically, do not neglect group velocity dispersion altogether, guaranteeing accurate modeling of the phase-matching processes \cite{JankowskiFejer:24}. At last, we neglect the effects of self- and cross-phase-modulation effectuated by the signal and idler fields. A more detailed analysis of these approximations is presented in Appendix A.

\section{III. The modeling framework}

\begin{table*}
    \caption{Simulation setup. Material parameters are adapted from \cite{Zelmon:97,Malitson:65,Nikogosyan:06,ShamsLoncar:22}, reference values for waveguide losses from \cite{ZhangLoncar:17,DesiatovLoncar:19,LukeZhang:20,KrasnokutskaPeruzzo:18}}
    \label{tbl:Table}
    \includegraphics[width=2\columnwidth]{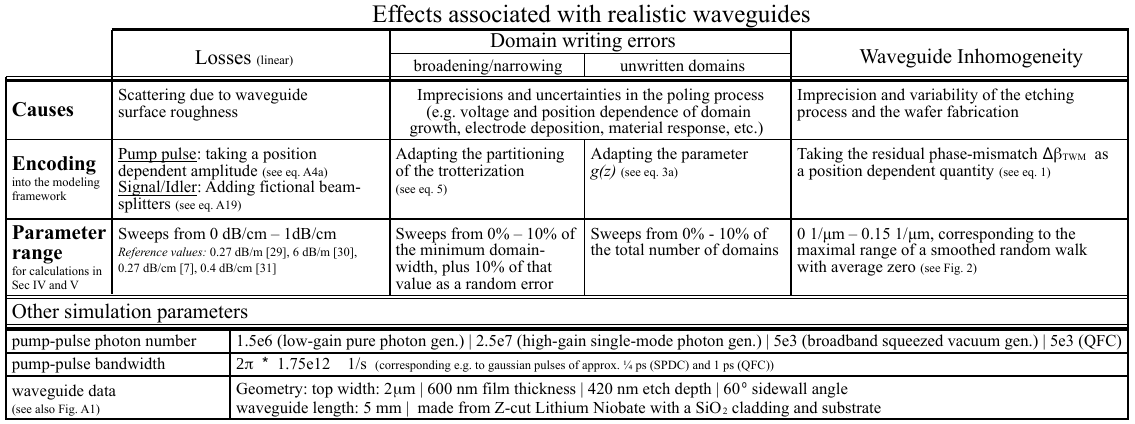}
\end{table*}

\begin{figure*}
    \centering
    \includegraphics[width=2.1\columnwidth]{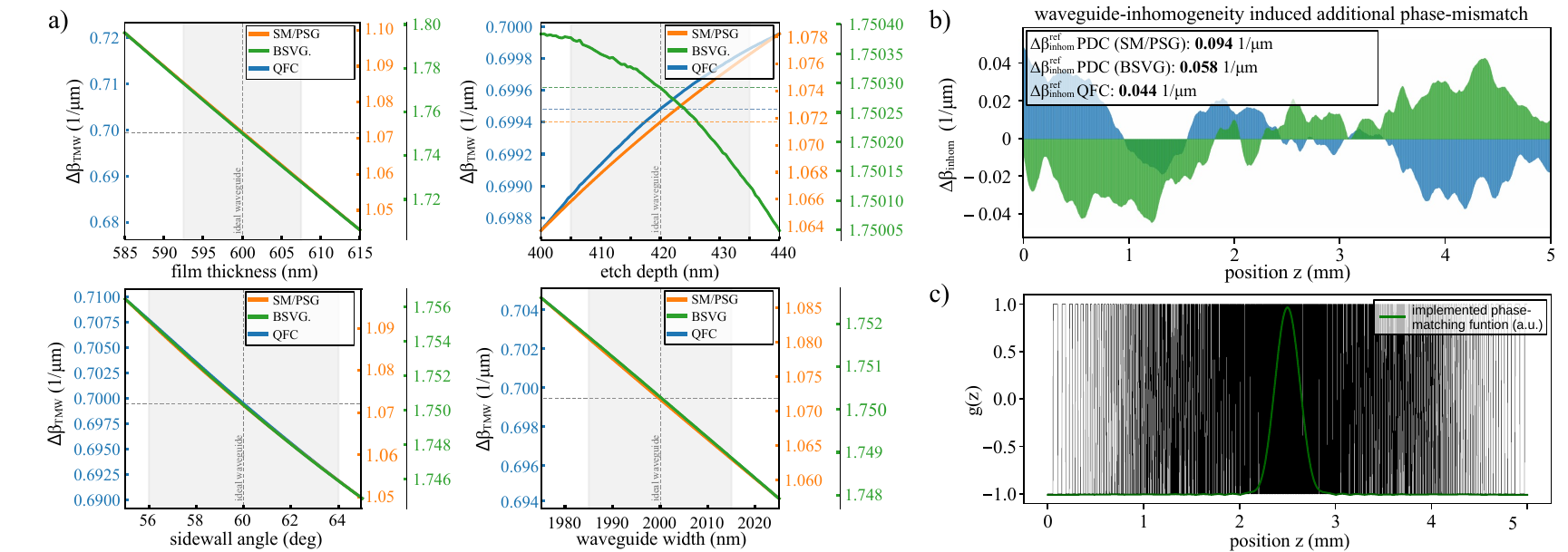}
    \caption{\textbf{a}) Effects of changes of different aspects of the waveguide geometry on the phase-mismatch of waveguides configured for single-mode/pure state generation (SM/PSG), broadband squeezed vacuum generation (BSVG), and QFC. We note, that, for every case, variations of the film thickness, a parameter rooted in the wafer fabrication, imparts the largest additional phase-mismatch. The dashed lines highlight the ideal, targeted waveguide geometry and the phase-mismatch that is compensated by the poling. \textbf{b}) Depiction of the parameter $\Delta \beta_{\text{\tiny TWM}} (z)$, used to capture the effects of waveguide inhomogeneities in our modeling framework (see Sec III). The data describing $\Delta \beta_{\text{\tiny TWM}} (z)$ is generated from a smoothed random walk with zero average and total range corresponding to the combination of geometry parameters in a) that give the maximal additional phase-mismatch. This range is quoted as 'maximal phase-mismatch' in Figs. \ref{fig:PPGTable}-\ref{fig:QFCTable}. The inset depicts reference values corresponding to the shaded areas in a), which, in our opinion, should be well attainable in practice. \textbf{c}) The poling pattern used to implement the gaussian phase-matching function used for single-mode/pure state generation and QFC. The pattern was calculated using \cite{AbranczykCustomPoling}.}
    \label{fig:WGInhomogeneity}
\end{figure*}

To model quantum nonlinear parametric interaction under the constraints discussed above, we follow a modeling framework developed by Quesada and coworkers \cite{QuesadaSipe:20,HeltQuesada:20,QuesadaSipe:22}, extend it to QFC and adapt it to account for the effects of waveguide inhomogeneity, losses and imperfect nonlinear domain engineering that are associated with realistic waveguides. We sketch this approach in the following section, present justifications of the respective approximations together with an outline of its derivation in Appendix A and B, and would like to refer to the high-quality original works for an in-depth, bottom-up treatment.

Our framework is based on a Heisenberg-picture propagator which connects the input (vacuum) fields at the beginning of the waveguide with those at its end, capturing both the linear propagation and the nonlinear interaction between the respective fields. The corresponding equations of motion (EOM) relate the spatial evolution of frequency-space signal and idler creation/annihilation operators. For PDC and QFC processes affected by pump-induced self- and cross-phase modulation, these equation take the form:
\begin{subequations}
\label{eq:BasicEOM}
\begin{align}
    \begin{split}
        {}\frac{\partial}{\partial z} &a_{s} (z,\omega) \: = \: i \Delta k_{s}(\omega) a_{s}(z,\omega) \: + \: i \frac{C_{\text{\tiny TWM}}^{\orange{*}} \: g(z)}{\sqrt{2 \pi}} \\
        &\times \: e^{\cpm i\int \Delta k_{\text{\tiny TWM}} (z) \mathrm{d}z} \int \mathrm{d} \omega' \mathcal{A}^{\orange{*}}(z,\orange{-} \omega + \omega') a_{i}^{\blue{\dagger}}(z,\omega') \\ 
        & + \: i \frac{C_{\text{\tiny XPM}, s} \: h(z)}{\sqrt{2 \pi}} \int \mathrm{d} \omega' \mathcal{E}(\omega - \omega') a_{s}(z,\omega') 
    \end{split}\\
    \begin{split}
        \frac{\partial}{\partial z} &a_{i}^{\blue{\dagger}} (z,\omega) \: = \: \cmp i \Delta k_{i}(\omega) a_{i}^{\blue{\dagger}}(z,\omega) \: \cmp \: i \frac{C_{\text{\tiny TWM}}^{\blue{*}} \: g(z)}{\sqrt{2 \pi}} \\
        &\times \: e^{\cmp i\int \Delta \beta_{\text{\tiny TWM}} (z) \mathrm{d}z} \int \mathrm{d} \omega' \mathcal{A}^{\blue{*}}(z,\omega \cpm \omega') a_{s}(z,\omega') \\ 
        & \cmp \: i \frac{C_{\text{\tiny XPM}, i} \: h(z)}{\sqrt{2 \pi}} \int \mathrm{d} \omega' \mathcal{E}(\omega - \omega') a_{i}^{\blue{\dagger}}(z,\omega') 
    \end{split}
\end{align}
\end{subequations}
wherein the annotations in blue and orange correspond to PDC and QFC processes respectively. The EOM include in particular: the terms $\Delta k_{s,i}(\omega)$ relating the walk-off between the pump and signal/idler fields, nonlinear interaction coefficients $C$ including waveguide field-overlap integrals, the pump-pulse amplitude $\mathcal{A}(z,\omega)$ (including effects of self-phase modulation) and energy distribution $\mathcal{E}(\omega)$, factors $h(z)$ and $g(z)$ indicating the presence and the sign of the material nonlinearity, and a potential residual phase-mismatch $\Delta \beta_{\text{\tiny TWM}} (z)$, corresponding to the respective three-wave-mixing processes. We give explicit expressions of these quantities in Appendix B.

To perform calculations, we discretize the operators $a_{j}(z,\omega)$ for a set of $N$ frequencies within signal and idler frequency bands of interest $\omega_{j,n} = \omega_{j,0} + n \Delta \omega |_{n=1}^{N}$ and write them as column vectors as ${\bf a}_{j}(\omega) = \left[ a_{j}(z,\omega_{1}), ..., a_{j}(z,\omega_{N}) \right]^{\mathrm{T}}$, with $j \in \{s,i\}$. This allows to write the discretized EOM
\begin{equation}
    \label{eq:DiscretizedEOM}
    \frac{\partial}{\partial z} 
        \begin{pmatrix}
        {\bf a}_{s}(z) \\
        {\bf a}_{i}^{\blue{\dagger}}(z)
        \end{pmatrix}
    \: = \: i \underbrace{
        \left[ \begin{array}{c|c}
        {\bf G}(z) & {\bf F}^{\orange{*}}(z) \\
        \hline
        \blue{-} \left[ {\bf F}^{\mathrm{T}} (z) \right]^{\blue{*}} & \; \blue{-} {\bf H} (z) \; \; \:\\
        \end{array} \right]
        }_{\coloneq {\bf Q}(z)}
        \begin{pmatrix}
        {\bf a}_{s}(z) \\
        {\bf a}_{i}^{\blue{\dagger}}(z)
        \end{pmatrix}
        ,
\end{equation}
where again annotations in blue and orange refer to models of PDC and QFC respectively. The entries of the matrices ${\bf F}(z)$, ${\bf G}(z)$ and ${\bf H}(z)$ correspond to the coefficients in the EOM (eq. \ref{eq:BasicEOM}) at the respective frequencies:
\begin{subequations}
\begin{align}
    \begin{split}
        {\bf F}_{mn}(z) = \: &\frac{C_{\text{\tiny TWM}} \: g(z)}{\sqrt{2 \pi}} e^{i\int \Delta \beta_{\text{\tiny TWM}} (z) \mathrm{d}z} \\
        &\times \: \mathcal{A}(z,\omega_{i,n} \cpm \omega_{s,m}) \Delta \omega
    \end{split}\\
    \begin{split}
        {\bf G}_{mn}(z) = \: &\Delta k_{s}(\omega_{s,m}) \delta_{m,n} \: \\
        &+ \: \frac{C_{\text{\tiny XPM}, s} \: h(z)}{2 \pi} \mathcal{E}(\omega_{s,m} - \omega_{s,n}) \Delta \omega
    \end{split}\\
    \begin{split}
        {\bf H}_{mn}(z) = \: &\Delta k_{i}(\omega_{i,m}) \delta_{m,n} \: \\
        &+ \: \frac{C_{\text{\tiny XPM}, i} \: h(z)}{2 \pi} \mathcal{E}(\omega_{i,m} - \omega_{i,n}) \Delta \omega
    \end{split}
\end{align}
\end{subequations}
It is now straight-forward to solve the EOM by by formally integrating eq. (\ref{eq:DiscretizedEOM}) 
\begin{equation}
\label{eq:PropagatorEOM}
    \begin{aligned} 
            \begin{pmatrix}
            {\bf a}_{s}(z) \\
            {\bf a}_{i}^{\blue{\dagger}}(z)
            \end{pmatrix}
        &= \: {\bf K}(z,z_{0})
            \begin{pmatrix}
            {\bf a}_{s}(z_{0}) \\
            {\bf a}_{i}^{\blue{\dagger}}(z_{0})
            \end{pmatrix} \\
        &= \:   
            \left[ \begin{array}{c|c}
            {\bf K}^{s,s}(z,z_{0}) & \left[ {\bf K}^{s,i}(z,z_{0}) \right] \\
            \hline
            \left[ {\bf K}^{i,s}(z,z_{0}) \right]^{\blue{*}} & \left[{\bf K}^{i,i}(z,z_{0}) \right]^{\blue{*}} \\
            \end{array} \right]
            \begin{pmatrix}
            {\bf a}_{s}(z_{0}) \\
            {\bf a}_{i}^{\blue{\dagger}}(z_{0})
            \end{pmatrix},
    \end{aligned}
\end{equation}
which is now fully characterized by the propagator ${\bf K}$, calculated from the Trotter-Suzuki expansion 
\begin{equation}
\label{eq:DefinitionPropagator}
    {\bf K}(z,z_{0}) \: = \: \prod_{l=1} \mathrm{exp} \left[ i {\bf Q}(z_{l}) \Delta z_{l} \right] 
\end{equation}
over intervals $\Delta z_{l}$ in which ${\bf Q}(z_{l})$ is constant; $z_{l} = z_{0} + \sum_{l'=1}^{l} \Delta z_{l'}$ and $\sum_{l} \Delta z_{l} = z - z_{0}$.

The propagator ${\bf K}$ then allows calculation of the joint spectral amplitude (JSA) of the PDC process at the discretized frequencies $\omega_{s/i,n}$ 
\begin{equation}
    {\bf J} = {\bf V} \left[ \mathrm{\textbf{diag}} \Bigl( \underbrace{ \frac{1}{2} \mathrm{sinh}^{-1} \left( 2 \tilde{r}_{1} \right) }_{\coloneq r_{1}} ,...,\underbrace{ \frac{1}{2} \mathrm{sinh}^{-1} \left( 2 \tilde{r}_{N} \right) }_{\coloneq r_{N}} \Bigr) \right] {\bf W}^{\mathrm{T}},
\end{equation}
wherein ${\bf V}$, ${\bf W}$ and the diagonal matrix ${\bf D}$ with entries ${\bf D} = \mathrm{\textbf{diag}}\left( \tilde{r}_{1}, ..., \tilde{r}_{N} \right)$ represent the singular value decomposition of the phase sensitive moment ${\bf M}$, which is calculated from the propagator as ${\bf M} = {\bf K}^{s,s}(z,z_{0}) \left[ {\bf K}^{i,s}(z,z_{0}) \right]^{\mathrm{T}} = {\bf V} {\bf D} {\bf W}^{\mathrm{T}}$. The $r_{n}$ correspond to the squeezing parameters of the respective Schmidt modes. The cross-coupling of the signal and idler modes of the QFC process, similarly, is fully captured by the matrices ${\bf K}^{s,i}$ and ${\bf K}^{i,s}$.

As we show in Appendix B, the propagator ${\bf K}$ further allows to directly write the states after interaction, calculate the down-converted photon number $\mathcal{N}_{s,i}$, the conversion probability $\gamma$ of the QFC process, and spectral and entanglement characteristics in terms of input and output Schmidt modes.

We note, that describing the interaction using the propagator ${\bf K}$, does not require introducing a cutoff of the Fock-space or the Schmidt-modes.

To end this section, we present a short summary of the figures-of-merit used to evaluate the calculations presented in Secs. IV and V:

\textbf{Figures of merit:} Quantitatively comparing the interaction with and without the different effects associated with realistic waveguides requires appropriate figures-of-merit. In the following, we will rely on the down-converted photon number and the conversion probability as indicators of changes in the strength of the interaction. We further use the Schmidt number (defined as $\mathrm{SN} = \big[ \sum_{j=1}^{N} \mathrm{sinh}^{2}(r_{j}) \big]^{2} / \sum_{j=1}^{N} \mathrm{sinh}^{4}(r_{j})$, and related to the heralded spectral purity as $\mathcal{P} = 1 / \mathrm{SN}$) to relate the modal structure of the states generated via PDC. As for QFC, the purity of the output depends on the input state as well as the interaction, we quantify the device via the separability of its dominant Schmidt mode (defined as $\mathrm{S} = \gamma / \sum_{j=1}^{N} \gamma$) pertaining to its relative conversion efficiency. Where applicable, we further consider the squeezing parameter $r_{1}$ of the dominant Schmidt mode.


\section{IV. Nonlinear parametric interaction in realistic waveguides}

\begin{figure*}
    \centering
    \includegraphics[width=1.8\columnwidth]{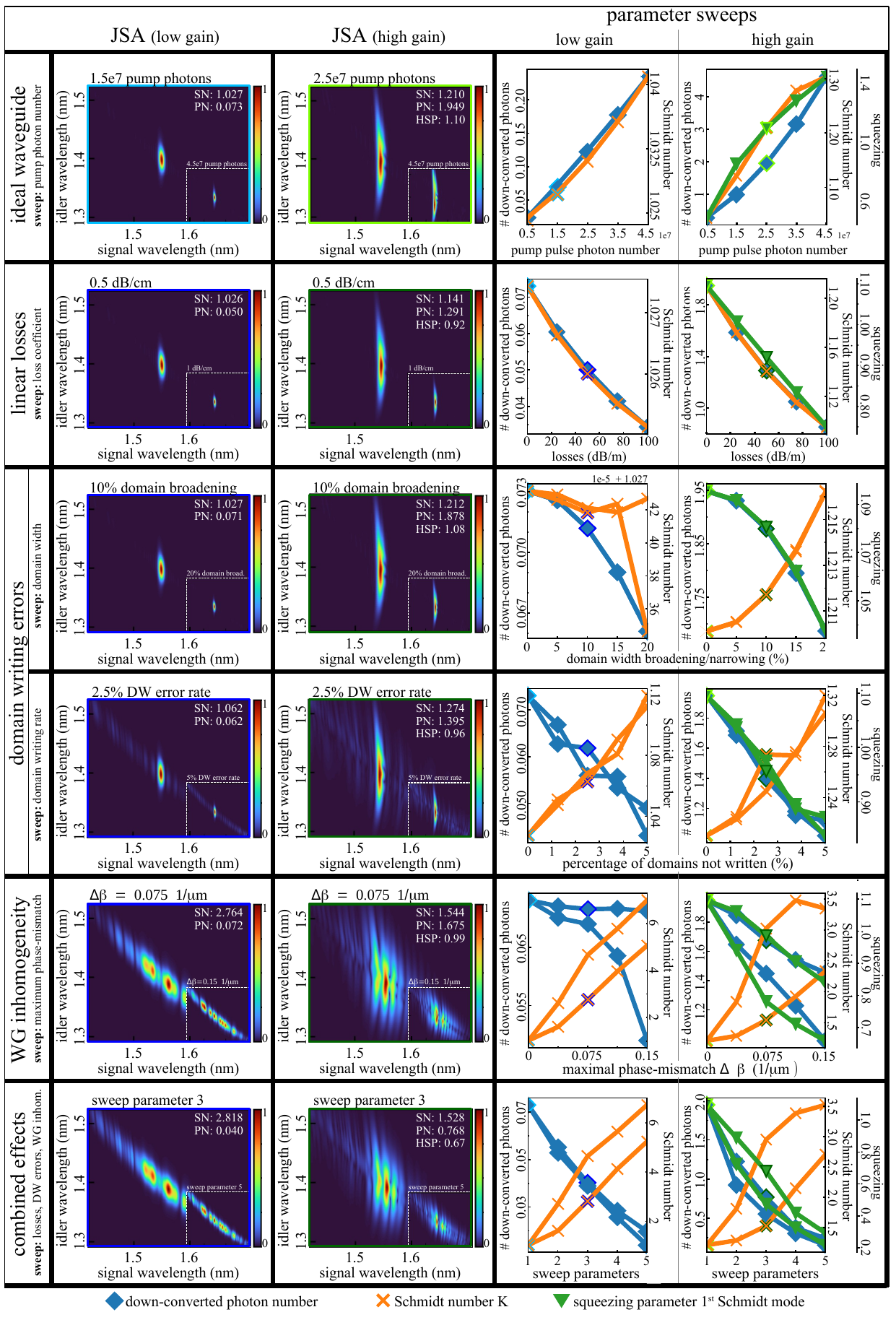}
    \caption{\textbf{Single-mode/pure state generation:} JSAs in the left columns corresponds to markers highlighted in identical color in the parameter sweeps depicted in the right columns.  Markers highlighted in light-blue and -green correspond to perfect waveguides with pump-pulses of 1.5e6 and 2.5e7 photons respectively. SN: Schmidt-number, PN: Number of down-converted photons, HSP: Squeezing parameter of the dominant Schmidt-mode, WG: Waveguide, DW: Domain writing.}
    \label{fig:PPGTable}
\end{figure*}

\begin{figure*}
    \centering
    \includegraphics[width=1.8\columnwidth]{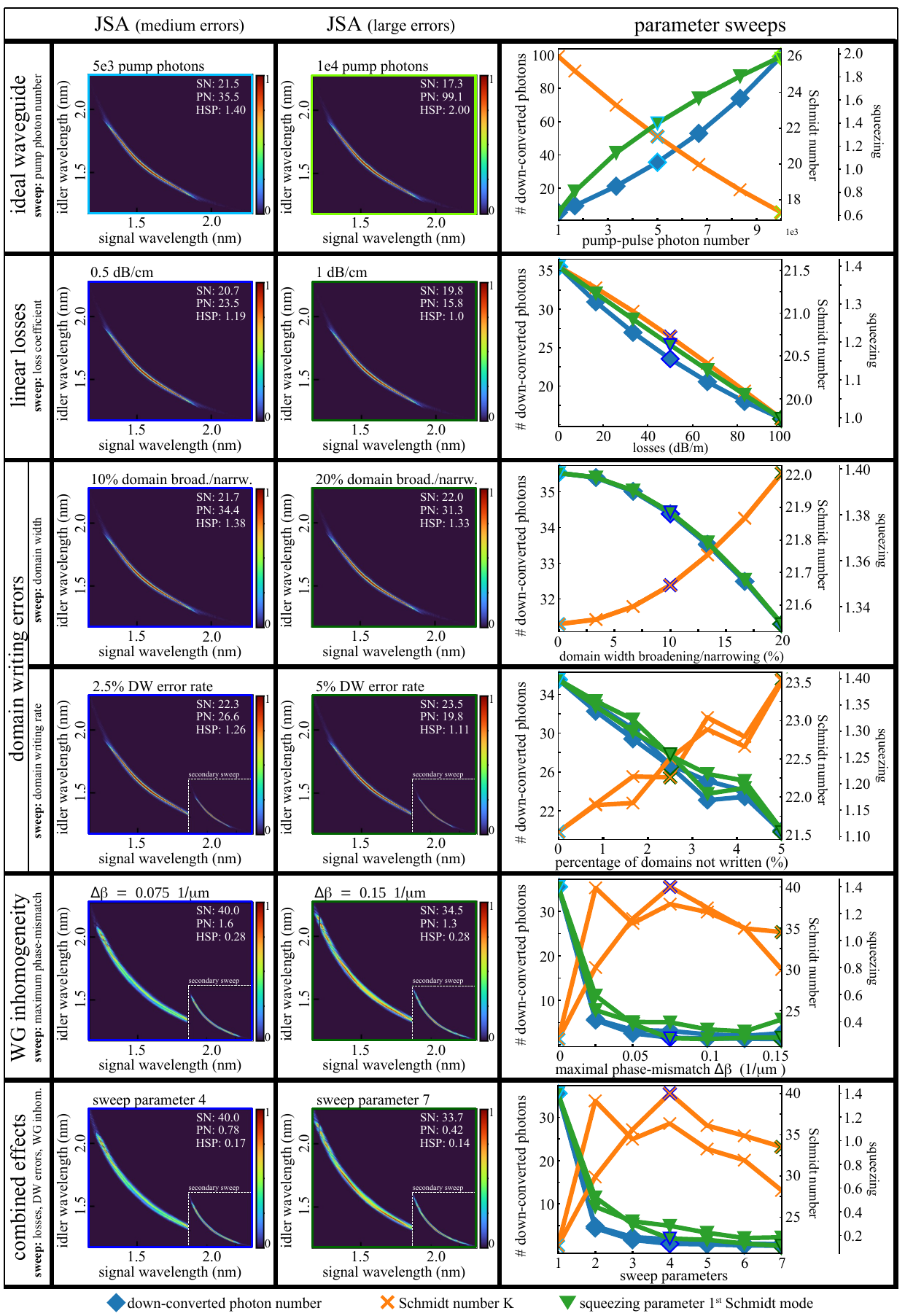}
    \caption{\textbf{Broadband squeezed vacuum generation:} JSAs in the left columns corresponds to markers highlighted in identical color in the parameter sweep depicted in the right column. Markers highlighted in light-blue correspond to perfect waveguides with pump-pulses of 5e3 photons. The curvature of the generated spectrum is a result of plotting in terms of wavelengths. SN: Schmidt-number, PN: Number of down-converted photons, HSP: Squeezing parameter of the dominant Schmidt-mode, WG: Waveguide, DW: Domain writing.}
    \label{fig:BSQTable}
\end{figure*}

\begin{figure*}
    \centering
    \includegraphics[width=1.8\columnwidth]{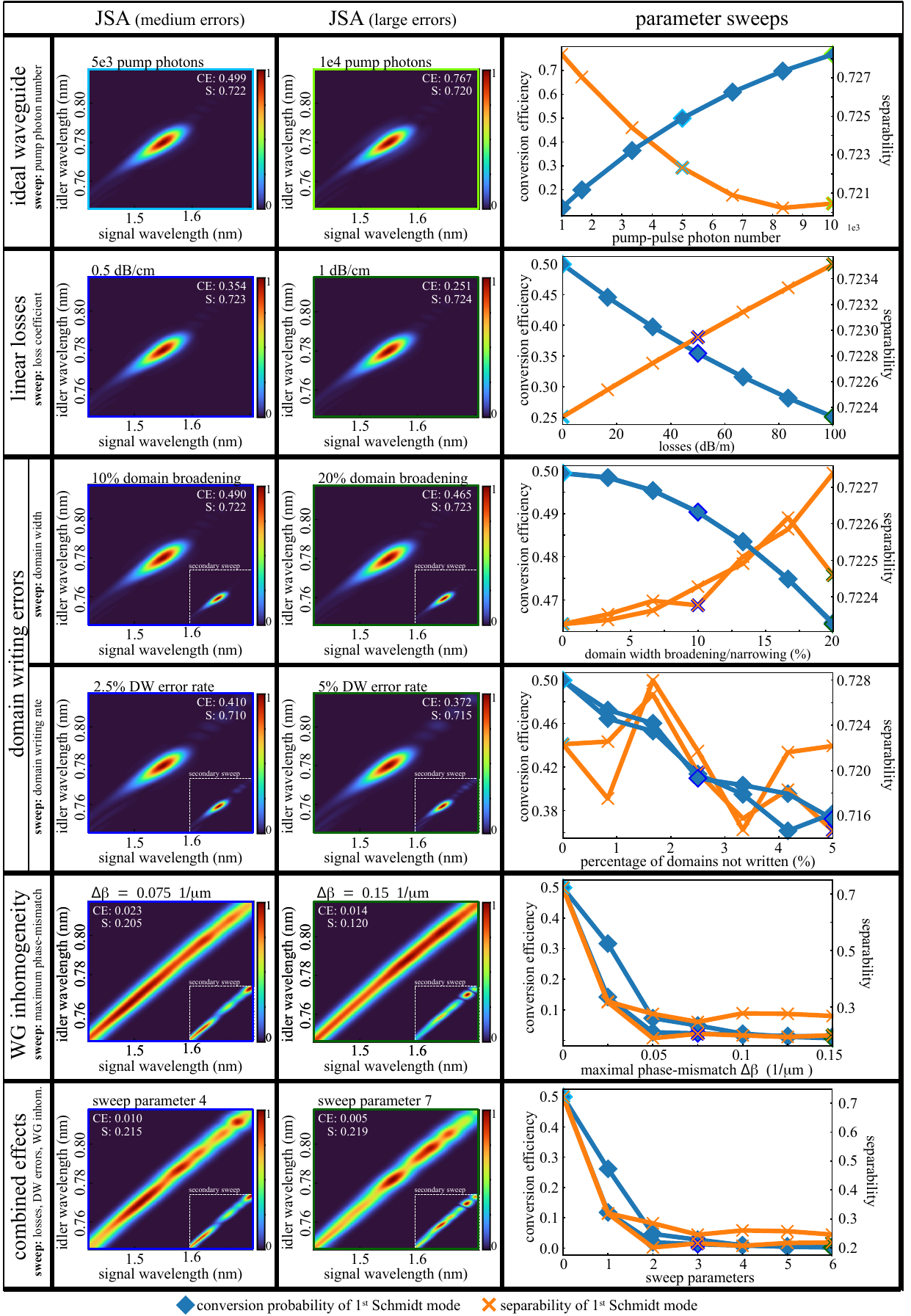}
    \caption{\textbf{Quantum frequency conversion:} JSAs in the left columns corresponds to markers highlighted in identical color in the parameter sweep depicted in the right column. Markers highlighted in light-blue correspond to perfect waveguides with pump-pulses of 5e3 photons. CE: Conversion efficiency, S: Separability, WG: Waveguide, DW: Domain writing.}
    \label{fig:QFCTable}
\end{figure*}

To quantitatively analyze the states generated from nonlinear interaction in realistic waveguides, we consider an exemplary device consisting of a single nonlinear waveguide configured for either heralded pure-state/single-mode squeezed-state generation, broadband squeezed vacuum generation, or QFC, under each of the effects described in Sec. II and depicted in Fig. \ref{fig:ConceptualFig}.  Table \ref{tbl:Table} gives an overview of these effects, their encoding into the the modeling framework described Sec III, and the range of parameters considered in our analysis. An illustration of the derivation and the treatment of the waveguide inhomogeneity is presented in Fig. \ref{fig:WGInhomogeneity}.

To perform calculations, we explicitly specify a number of experimental parameters - like, for instance, the interacting wavelengths and spatial modes, the waveguide and the pump-pulse data - and select them to be representative of the most common examples of waveguided nonlinear interaction \cite{NehraMarandi:22,WuDiddams:24,KashiwazakiFurusawa:20,LuTang:19}. As argued in Appendix A, we expect qualitative deviations from this exemplary setting only for extreme cases operating at the edge of their functional range. To highlight and isolate the effects associated with realistic waveguides, we perform the majority of calculations in this section below the threshold of significant self-phase modulations; a study of the impact of pump induced self- and cross-phase modulation is presented separately in Sec V.

We present a detailed analysis of the impact of scattering losses, domain writing errors, and waveguide inhomogeneity for a nonlinear waveguide operated to generate heralded pure photons and single-mode squeezed states in the low- and high-gain regimes, broadband squeezed vacuum and QFC in Figs. \ref{fig:PPGTable}-\ref{fig:QFCTable} respectively. Specifically, we calculate the number of down-converted photons (or, for QFC, the conversion probability), the Schmidt number pertaining to the state purity, and, where applicable, the squeezing parameter of the dominant Schmidt mode, for a selection of parameters ranging from a perfect to a highly erroneous waveguide (see Table \ref{tbl:Table}). Figs. \ref{fig:PPGTable}-\ref{fig:QFCTable} further depict exemplary JSAs for a selection of these parameters, sweeps of the cumulative impact of the different effects and, assuming a perfect waveguide, the pump photon number. We note, that the different figures of merit correspond solely to the frequency ranges depicted in the respective JSAs, and do not take into account any potential interaction outside of this range.

We give, in the following, a short, qualitative summary of our observations:

\textbf{Linear losses:} As would be expected, increasing waveguide losses lead to a decrease of the conversion efficiency and the associated squeezing. Purity and separability figures-of-merit increase slightly with increased losses, an effect explored more closely in \cite{HoudeQuesada:23}. Interestingly, for broadband squeezed vacuum generation, the state purity increases with increasing pump photon numbers.

\textbf{Domain writing errors:} Domain broadening and narrowing are functionally identical, and have comparatively little impact on device operation, effectuating only minor performance decreases. Effects of not-written domains are more pronounced, leading to noticeable changes in conversion efficiency and state purity/separability, which, however, should be well manageable for realistic error rates.

\textbf{Waveguide Inhomogeneity:} Variations of the waveguide geometry, and thus the associated phase-mismatch $\Delta \beta_{\text{\tiny TWM}} (z)$, considerably affect the device performance, both significantly decreasing the conversion efficiency and smearing the interaction across a broad range of frequencies, substantially decreasing the purity/separability of the generated states. There appears to be a considerable variability to these effects, indicating sensitivity to how the geometry varies, in addition to its magnitude. This indicates a large unpredictability of the impact of this effect on an actual device.

Qualitatively, all processes appear to be affected similarly. In application, effects impacting the conversion efficiency much stronger than the state purity (i.e. losses and, arguably, domain broadening/narrowing), are especially problematic when employed for the generation of squeezed states, but may be tolerable when used for single-mode/pure state generation. The apparently sharp drop-off of the figures-of-merit in the plots depicting the impact of waveguide-inhomogeneities observed for broadband squeezed vacuum generation and QFC is likely due to a significant part of the interaction occurring outside of the frequency window considered here.

\begin{figure*}[b]
    \centering
    \includegraphics[width=2\columnwidth]{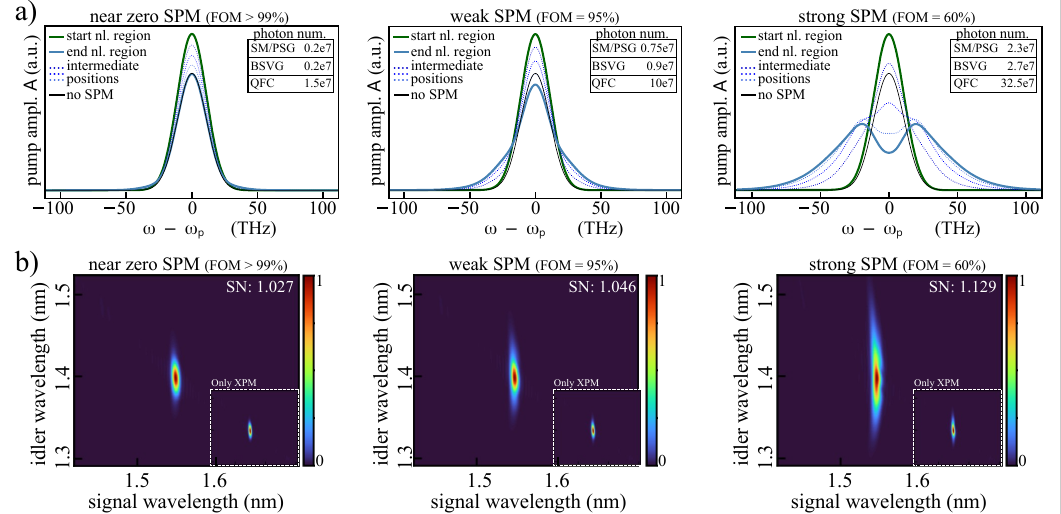}
    \caption{\textbf{a}) Effect of self-phase modulations on the pump spectral amplitude $\mathcal{A}$ (see eq. \ref{eq:PumpSpectralAmp}). To quantify regimes with near-zero, weak and strong self-phase modulation we define the overlap between the self-phase modulated pump amplitude (solid blue line) and a pump amplitude evolving without self-phase modulation (black line), at the end of the waveguide, as a figure-of-merit (FOM). The inset tables give the breaking points, in terms of the pump-photon number, of the respective regimes. The difference of these values is rooted in the size of the modal fields of the respective pump fields, which are smaller for shorter wavelengths. \textbf{b}) JSAs of the states generated under no, weak and strong self- and cross-phase modulation (as defined in a)) for the generation of heralded pure states or single-mode squeezed-states. As the impact of cross-phase modulation (XPM) is somewhat harder to quantify than that of self-phase modulation (SPM), we give JSA's affected solely by cross-phase modulation in the insets. All plots assume linear losses of $0.5 \, \mathrm{dB/cm}$. SM/PSG: Single-mode/pure state generation, BSVG: Bright squeezed vacuum generation, SN: Schmidt number, nl: Nonlinear.}
    \label{fig:SXPM}
\end{figure*}

\section{V. Impact of self- and cross-phase modulation}
For sufficiently strong nonlinear interaction, self- and cross-phase modulation effectuated by the pump field begin to significantly affect the states generated from both PDC and QFC. The interaction mediating self- and cross-phase modulation is, unlike those responsible for PDC and QFC processes, unaffected by efficiency reductions associated with artificial phase-matching (periodic or engineered), and, in the case of self-phase modulation, further independent of the modal overlap between the pump, signal and idler fields. 

Accordingly, the point at which these effects become considerable - in terms of the gain of PDC or QFC - can depend significantly on both the type of interaction and how it is configured. In Fig. \ref{fig:SXPM} we explore the onset of significant self- and cross-phase modulation by scaling the pump power - related directly to the strength of the nonlinear interaction - and its impact on the spectral properties of the generated states. We find, in particular, that due to the gaussian domain engineering and the reduced modal overlap necessary to satisfy asymmetric GVM, the maximal squeezing available from PDC configured for single-mode/pure state generation is ultimately limited by the self-phase modulation of the pump field, at squeezing parameters much below those of PDC configured for the generation of broadband squeezed vacuum.

The gain associated with the onset of significant self- and cross-phase modulation for the configurations mediating broadband squeezed vacuum generation and QFC (as  they are considered in the previous section, both optimized for nonlinear efficiency) occur beyond the validity of the parametric approximation even if linear losses are considered, and can not be captured within our model. We note, that this indicates either the experimental feasibility of implementing single-pass nonlinear interaction strong enough to deplete the pump field - marking the onset of non-gaussian quantum optics at relatively low input powers - or point toward fabrication errors like to ones discussed here, as the reason to why it is not.  

\section{VI. Discussion and Outlook}
In this paper, we presented a modeling framework that captures realistic quantum nonlinear interactions in waveguided systems by explicitly integrating the engineering methods used for designing integrated waveguides with the quantum mechanical theory of nonlinear parametric interaction. We believe that this approach will facilitate the design of higher-quality and more efficient nonlinear devices, representing a key step toward the technological scaling of quantum light sources.

Leveraging these capabilities, we conducted a comprehensive study of how common waveguide fabrication errors affect the quantum states generated through nonlinear interaction. Our analysis quantifies the critical role of high-quality fabrication and underscores the need for accurate modeling tools to characterize these imperfections. In particular, the study reveals the significance of including complex deviations from ideal designs---most notably variations in waveguide geometry, which have previously been considered primarily in the context of classical interactions \cite{ChenFan:24}---in addition to waveguide losses. We find that inhomogeneities in the waveguide structure are especially sensitive to variations in thin-film thickness, which originate during wafer fabrication and are typically beyond the control of most research laboratories. We note that explicit assessments of current waveguide fabrication standards could refine the parameter ranges presented in Table \ref{tbl:Table} and substantially enhance the accuracy of future modeling efforts.

We further quantitatively investigated the onset of critical pump-induced self- and cross-phase modulation, identifying it as a key limiting factor in the generation of single-mode squeezed states using asymmetric group velocity matching. This limitation arises from the reduced down-conversion efficiency associated with such configurations. Overcoming this challenge---whether through alternative strategies for single-mode state generation or through the development of advanced material platforms with more favorable dispersion properties \cite{MccrackenFedrizzi:18}---would represent a significant advancement in the field.

In principle, the modeling framework presented in this work can be readily extended to more complex photonic structures—both linear and nonlinear \cite{KimSohn:24}---ultimately enabling the precise simulation of intricate nonlinear quantum photonic circuits. In practice, however, such ambitions are often constrained by the growing complexity of the underlying models. Reducing this complexity---through adaptations of existing theoretical methods to lower computational overhead, as well as by developing circuit-specific models that focus only on relevant parameters---constitutes a critical direction for future research. Another promising avenue involves incorporating the modeling of measurement processes, which naturally raises important questions about what aspects of the system must be modeled in the first place.

Strong nonlinear interactions, such as those encountered in the high-gain regime of broadband squeezed vacuum generation, are not adequately captured by the parametric approximation. This highlights the need for models that account for pump depletion and, ideally, extend beyond the Gaussian framework of quantum optics \cite{YanagimotoMabuchi:24, JankowskiFejer:24}. Reapplying quantitative modeling to realistic waveguides within a framework suitable for non-Gaussian quantum optics may represent a critical step toward the experimental generation and characterization of non-Gaussian states. Similarly, extending such analysis to nonlinear interactions in waveguide resonators or those driven by continuous-wave lasers could offer valuable new insights.

Overcoming the limitations on single-mode squeezed-state generation imposed by pump self-phase modulation, potentially through pump pulse shaping, temporal trapping techniques \cite{YanagimotoHamerly:22}, or advanced dispersion engineering to enable more efficient group velocity matching \cite{LucasPapp:23, GuoKippenberg:20}, would be of considerable interest. Extending the modeling framework to capture nonlinear interactions mediated solely by third-order nonlinearities, or exploring regimes where self-phase modulation is significant or squeezing leads to appreciable pump depletion, represents another promising direction. Additionally, incorporating nonlinear loss mechanisms into the model could provide a more comprehensive understanding of realistic system dynamics.

\smallskip
\noindent\textbf{Author contributions}\\
T.W. conceived the project, implemented the modelling framework and generated the results. A.Y. assisted with supercomputer implementation. A.P. supervised the project, T.W. and A.P. wrote the paper.

\smallskip
\noindent
\textbf{Acknowledgments}\\
A.P. acknowledges an RMIT University Vice-Chancellor’s Senior Research Fellowship and a Google Faculty Research Award. This work was supported by the Australian Government through the Australian Research Council under the Centre of Excellence scheme (No: CE170100012).

\smallskip
\noindent\textbf{Data availability}\\
All data generated or analyzed during this study are included in this published article

\smallskip
\noindent\textbf{Competing interests}\\
The authors declare no competing interests.

\appendix
\renewcommand{\thefigure}{A\arabic{figure}}
\setcounter{figure}{0}
\renewcommand{\theequation}{A\arabic{equation}}
\setcounter{equation}{0}

\section{Appendix A: Taking approximations}

\begin{figure*}
    \centering
    \includegraphics[width=2\columnwidth]{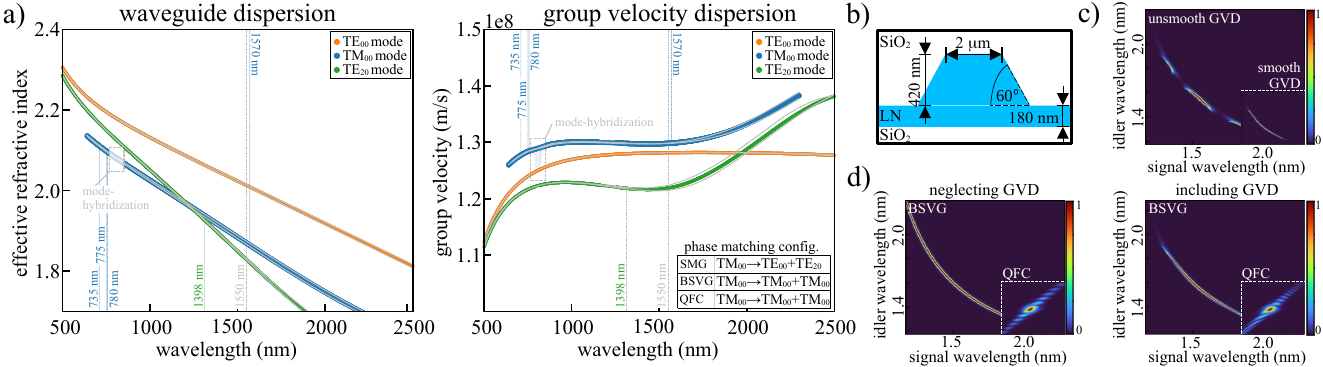}
    \caption{\textbf{a}) Dispersion properties of the waveguide depicted in \textbf{b}), considered for the quantitative analysis in Sections IV and V. The inset in the right figure depicts the modes selected for phase-matching. Phase-matching between higher order modes is necessary to achieve asymmetric GVM required for the generation of states with a single spectral-temporal mode (SMG); phase-matching between $\mathrm{TM_{00}}$ modes, as considered for broadband squeezed vacuum generation (BSVG) and QFC, yields the highest nonlinear interaction. The gray lines superimposed in the dispersion curves correspond to the variation of the dispersion curves under changes of the waveguide geometry, given by the reference cases of Fig. \ref{fig:WGInhomogeneity}a. \textbf{c}) We note that discontinuities in the refractive index (due to modal hybridization, see the box highlighted in gray) result in highly out-of-order group velocities, which can produce artifacts when calculating the nonlinear interaction in this region. As these out-of-order group velocities, to the best of our knowledge, do not describe physical reality, and the corresponding artifacts are not observed experimentally \cite{NehraMarandi:22}, we assume a smooth group velocity dispersion in this frequency range, but note, that the hybridization may lead to additional losses due to cross-modal coupling. \textbf{d}) Illustration of the importance of explicitly including group-velocity dispersion into the modeling framework: While in certain narrowband cases, group-velocity dispersion can be neglected with good approximation, for the configurations generating broadband squeezed vacuum and mediating QFC, there are critical differences. LN: Lithium Niobate.}
    \label{fig:AppdxA_GV}
\end{figure*}

\begin{figure}
    \centering
    \includegraphics[width=1\columnwidth]{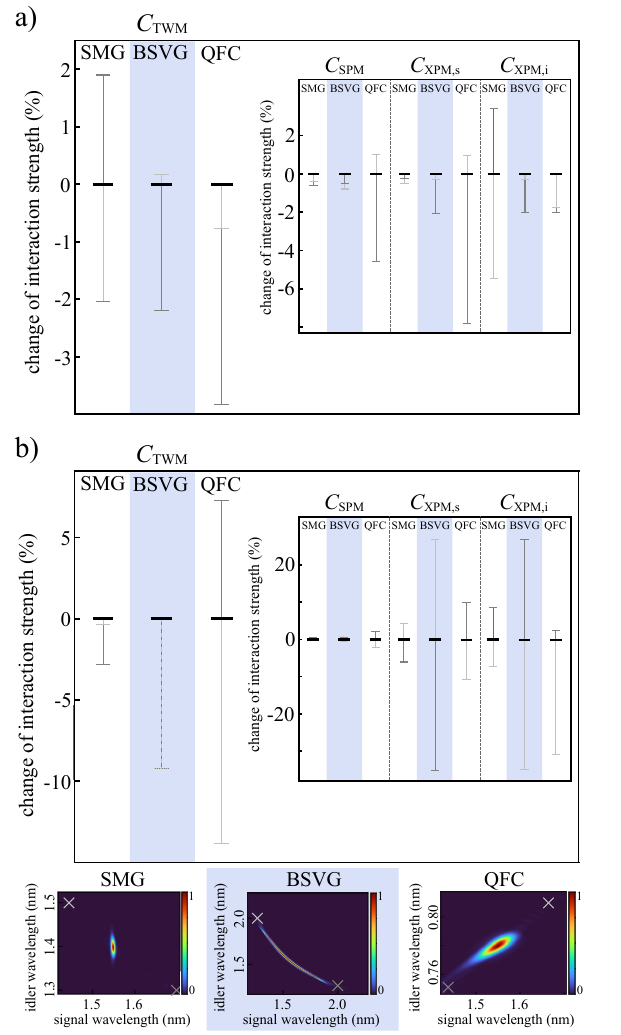}
    \caption{\textbf{a}) Variation of the different interaction coefficients $C$ (see eqs. \ref{eq:InteractionCoefficients}) under changes of the waveguide geometry for the reference values depicted in Fig. \ref{fig:WGInhomogeneity}b. \textbf{b}) Variation of the interaction coefficients $C$ under changes of the pump, signal and idler wavelengths. The considered signal and idler wavelengths are indicated in the joint-spectral-amplitudes depicted at the bottom; variations of $C_{\mathrm{SPM}}$ were calculated assuming deviations from the central frequency of the pump by the pulse bandwidth. We note, that if both edge cases yield a reduction of the interaction strengths, there may be intermediate values which give a minor increase. SMG: Single-mode/pure state generation, BSVG: Broadband squeezed vacuum generation.}
    \label{fig:AppdxA_IC}
\end{figure}

In the following, we present an analysis of the different approximations that are frequently employed when formally modeling quantum nonlinear interaction in integrated waveguides, together with a discussion of which are applicable to the model considered here. While the following study was carried out for an explicitly specified waveguide geometry, largely arbitrarily adopted from the literature, we note, that we expect the following analysis to be representative of and applicable to integrated waveguides in general, provided they confine the involved modes with sufficient quality.

\textbf{Group velocity dispersion:} The relative magnitude of the group velocities of the pump, signal and idler fields critically affect their nonlinear interaction, and, in waveguides, can be engineered through selection of the waveguide geometry and the interacting modes \cite{URenRaymer:06}. Accordingly, how group velocity dispersion is included in the modeling framework is crucial. We find, first, that variations of the waveguide geometry (see the gray lines in Fig. \ref{fig:AppdxA_GV}a, corresponding to the reference values in Fig. \ref{fig:WGInhomogeneity}b) start to noticeably affect the group velocity dispersion only at long wavelengths, where confinement becomes difficult. Consequently, we take the group velocity to be position independent for each of the processes considered here (see inset in Fig. \ref{fig:AppdxA_GV}a). We do, however, not neglect group velocity dispersion all together. While, in certain cases relating narrowband interactions, group velocity dispersion may be neglected in good approximation, including group velocity dispersion is critical when considering broadband interactions (see Fig. \ref{fig:AppdxA_GV}d). We note, that calculating the group velocity dispersion from an eigenmode solver can lead to out-of-order group velocities associated with hybridized waveguide modes, which can yield artifacts when calculating the interaction, and have to be corrected for.

\textbf{The interaction coefficients:} Our modeling frameworks takes the interaction coefficients $C$ (see eqs. \ref{eq:InteractionCoefficients}) to be independent of both the waveguide geometry (and thus the position $z$) and variations of the frequency within the pump, signal and idler frequency bands. We present a quantitative analysis of the impact of both parameters in Fig. \ref{fig:AppdxA_IC}. While variations of the waveguide geometry result in only a minor modification of the interaction coefficients (generally $<< 10\%$, see Fig. \ref{fig:AppdxA_IC}a), large variations of the signal and idler wavelengths have a more pronounced impact, particularly when considering broadband interactions (see Fig. \ref{fig:AppdxA_IC}a). Considering, however, that these variations corresponds to extreme cases, we believe that dispersion of the interaction coefficients within the pump, signal and idler frequency bands need only be considered when ultra-precise modeling is required, but are otherwise dwarfed by the effects discussed in Sec. IV.  

\textbf{Self- and cross phase modulation from signal and idler fields:} The strength of cross- and self-phase modulation scales directly with the strength of the field effectuating the interaction. Since signal and idler fields are orders-of-magnitude weaker than the pump field, self- and cross-phase modulation imparted by them becomes significant only much after the interaction is already dominated by pump induced self- and cross-phase modulation. Accordingly, for interactions operating at or below the onset of significant self- and cross-phase modulation effectuated by the pump field, the self- and cross-phase-modulation imparted by signal and idler fields can confidently be neglected.

\textbf{Cross-modal coupling:} Depending on the geometry of the waveguide, certain frequency ranges may feature waveguide modes with near identical propagation constants. Consequently, both modes satisfy the required phase-matching conditions, but, due to different group-velocities may evolve differently. This effect is highly localized in frequency space, and may be modeled by an additional frequency-local linear loss channel from the targeted waveguide mode. Due to the difficulty of quantifying such a loss channel our calculations do not include such considerations, but we note, that such an effect may affect the down-converted photon number in the frequency-range highlighted in gray in Fig. \ref{fig:AppdxA_GV}a.    
 
\section{Appendix B: Deriving the model}
\noindent
\textbf{Deriving the EOM:} We adopt a description of the electromagnetic field in terms of the quantized electric displacement field ${\bf D}$ \cite{BhatSipe:06}:
\begin{equation}
    \begin{aligned}
        {\bf D}_{p,s,i}({\bf r}) = &\int \mathrm{d}k \sqrt{\frac{\hbar \omega_{p,s,i}(k)}{4 \pi}} {\bf d}_{p,s,i}^{\perp}(x,y) e^{ikz} \\
        &\times b_{p,s,i}(k,t) + \mathrm{H.c.},
    \end{aligned}
\end{equation}
with creation and annihilation operators $b_{j}(k,t)$,$b_{j}^{\dagger}(k,t)$ satisfying the usual bosonic commutation relations $\bigl[ b_{j}(k), b_{j'}(k') \bigr] = 0$ and $\bigl[ b_{j}(k,t), b_{j'}^{\dagger}(k',t) \bigr] = \delta_{j,j'} \delta(k-k')$ with $j \in \{p,s,i\}$ (see refs. \cite{BhatSipe:06,QuesadaSipe:22}, for a description of the associated magnetic field). The transverse fields ${\bf d}_{p,s,i}^{\perp}(x,y)$ perpendicular to the propagation direction $z$ are calculated for an explicit waveguide structure, taken to be frequency independent within the pump, signal and idler frequency bands (see Appendix A), and are normalized according to \cite{QuesadaSipe:20,QuesadaSipe:22}
\begin{equation}
    \int \mathrm{d}x \mathrm{d}y \frac{{\bf d}_{p,s,i}^{\perp} \left[ {\bf d}_{p,s,i}^{\perp} \right]^{*}}{\epsilon_{0} n_{p,s,i}^{2}(x,y)} \frac{v_{ph}}{v_{g}} = 1;
\end{equation}
wherein $n_{j}$ corresponds to the material refractive index, and $v_{ph}= c /{n_{\mathrm{eff},j}}$ and $v_{g} = c / \bigl( n_{\mathrm{eff},j} + \omega \frac{\partial n_{\mathrm{eff},j}}{\partial \omega} \bigr)$ to the respective waveguide phase- and group-velocities.

To account for matrial and waveguide dispersion we expand the wavevector to first order around a central component $k_{j}$ as $k - k_{j} = (\omega - \omega_{j})/v_{g,j}(\omega_{j})$. For simplicity, we will write the waveguide group velocity as $v_{j}$ from here on; $j \in \{p,s,i\}$. 

We now introduce field operators $\varphi_{p,s,i}(z,t) = \int \frac{\mathrm{d}k}{\sqrt{2 \pi}} b_{p,s,i} (k,t) \mathrm{exp}\left[i(k-k_{p,s,i})z\right]$ describing the broadband, slowly varying spatial envelopes of the respective fields. This allows to write the linear and nonlinear three- and four-wave-mixing Hamiltonian as \cite{QuesadaSipe:20,QuesadaSipe:22}:
\small
\begin{subequations}
\label{eq:Hamiltonians}
\begin{align}
    \begin{split}
        H_{\mathrm{L}} = &\sum_{j=p,s,i} \hbar \omega_{j} \int \mathrm{d}z \varphi_{j}^{\dagger}(z,t) \varphi_{j}(z,t) + i \sum_{j=p,s,i} \frac{\hbar v_{j}}{2} \\
        &\times \int \mathrm{d}z \left[ \frac{\partial \varphi_{j}^{\dagger}(z,t)}{\partial z} \varphi_{j}(z,t) - \varphi_{j}^{\dagger}(z,t) \frac{\partial \varphi_{j}(z,t)}{\partial z} \right]
    \end{split}\\
    \begin{split}
        H_{\mathrm{NL}}& = - \hbar \int \mathrm{d}z \left[ \frac{1}{2} \tilde{C}_{\mathrm{\tiny SPM}}(z) \: \varphi_{p}^{\dagger}(z,t) \varphi_{p}^{\dagger}(z,t) \varphi_{p}(z,t) \varphi_{p}(z,t) \right. \\
        &+ \tilde{C}_{\mathrm{\tiny XPM,s}}(z) \: \varphi_{p}^{\dagger}(z,t) \varphi_{p}(z,t) \varphi_{s}^{\dagger}(z,t) \varphi_{s}(z,t) \\
        &+ \tilde{C}_{\mathrm{\tiny XPM,i}}(z) \: \varphi_{p}^{\dagger}(z,t) \varphi_{p}(z,t) \varphi_{i}^{\dagger}(z,t) \varphi_{i}(z,t) \\
        &+ \left. \left( \tilde{C}_{\mathrm{\tiny TWM}} (z) \: e^{i \Delta \beta_{\mathrm{\tiny TWM}} z} \varphi_{p}(z,t) \varphi_{s}^{\blue{\dagger}}(z,t) \varphi_{i}^{\dagger}(z,t) + \mathrm{H.c.} \right) \right]
    \end{split}
\end{align}
\end{subequations}
\normalsize
where we retain a potential, uncompensated phase-mismatch $\Delta \beta_{\text{\tiny TWM}} = \beta_{p} \cmp \beta_{s} - \beta_{i}$, now expressed in terms of waveguide propagation constants $\beta_{j}$. In the nonlinear Hamiltonian, effects of cross- and self modulation effectuated by the signal and idler modes are neglected due to the relative weakness of the fields. We then take the parametric approximation, expressing the pump as a from the nonlinear interaction undepleted, classical field ($\varphi_{p}(z) \rightarrow \langle \varphi_{p}(z) \rangle$), and write the pump, signal and idler fields in frequency space co-moving with the pump-pulse \cite{QuesadaSipe:20}:
\begin{subequations}
    \begin{align}
        \begin{split}
            \mathcal{A}(z,\omega) \: &= \: \sqrt{\hbar \omega_{p}} \: e^{i \frac{w - \omega_{p}}{v_{p}}z } \int \frac{\mathrm{d}t}{\sqrt{2 \pi / v_{p}}} \: e^{i \omega t} \langle \varphi_{p}(z,t) \rangle \\
            & = \: e^{i \omega t_{0}} \sqrt{\hbar \omega_{p}} \int \frac{\mathrm{d}z'}{\sqrt{2\pi v_{p}}} e^{-i \frac{\omega - \omega_{p}}{v_{p}}z'} \Lambda(z') \\
            &\; \; \; \; \; \times \: \mathrm{exp} \left[ i | \Lambda(z') |^{2} \int_{z'}^{z} \frac{C_{\mathrm{\tiny SPM}}(z'')}{v_{p}} \mathrm{d}z'' \right]
        \label{eq:PumpSpectralAmp}
        \end{split}\\
        \begin{split}
            a_{s,i}(z,\omega) \: &= \: e^{i \frac{\omega - \omega_{s,i}}{v_{p}}z} \int \frac{\mathrm{d}t}{\sqrt{2\pi/v_{s,i}}} e^{i \omega t } \varphi_{s,i}(z,t);
        \end{split}
    \end{align}
\end{subequations}
wherein $\Lambda(z)$ describes the (classical) pump-field amplitude, normalized to the total photon number ($\int | \Lambda(z) |^{2} \mathrm{d}z = \mathcal{N}_{p}$). The energy distribution of the pump field in the moving frame is given by
\begin{equation}
    \mathcal{E}(\omega) \: = \: e^{i\omega t_{0}} \hbar \omega_{p} \int \mathrm{d}z | \Lambda(z) |^{2} e^{-i \omega z / v_{p}}.
\end{equation}
With this, we can connect the coefficients $\tilde{C}$ of eq. (\ref{eq:Hamiltonians}), carrying the nonlinear overlap between the interacting waveguide modes, with the interaction coefficients $C$ in the main text as
\cite{QuesadaSipe:20,QuesadaSipe:22}:
\begin{widetext}
\begin{subequations}
\begin{align}
    \begin{split}
    C_{\mathrm{\tiny SPM}}(z) &= \frac{3}{\epsilon_{0}^{3} \hbar} \left( \frac{\hbar \omega_{p}}{2} \right)^{2} \int_{\chi^{(3)}} \mathrm{d}x \mathrm{d}y \frac{\chi_{jklm}^{(3)}(z)}{n_{j}^{2} n_{k}^{2} n_{l}^{2} n_{m}^{2}} \left[ d_{p}^{j}(x,y) \right]^{*} \left[ d_{p}^{k}(x,y) \right]^{*} \left[ d_{p}^{l}(x,y) \right] \left[ d_{p}^{m}(x,y) \right]
    \end{split}\\
    \begin{split}
    C_{\mathrm{\tiny XPM},s/i} \: h(z) &= \frac{\tilde{C}_{\mathrm{\tiny XPM}}(z)}{v_{p}v_{s/i} \hbar \omega_{p}} = \frac{3}{2 \epsilon_{0}^{3}} \frac{\omega_{s/i}}{v_{p}v_{s/i}} \int_{\chi^{(3)}} \mathrm{d}x \mathrm{d}y \frac{\chi_{jklm}^{(3)}(z)}{n_{j}^{2} n_{k}^{2} n_{l}^{2} n_{m}^{2}} \left[ d_{p}^{j}(x,y) \right]^{*} \bigl[ d_{s/i}^{k}(x,y) \bigr]^{*} \left[ d_{p}^{l}(x,y) \right] \bigl[ d_{s/i}^{m}(x,y) \bigr]
    \end{split}\\
    \begin{split}
    C_{\mathrm{\tiny TWM}} \: g(z) &= \frac{\tilde{C}_{\mathrm{\tiny TWM}}(z)}{\sqrt{v_{p}v_{s}v_{i} \hbar \omega_{p}}} = \frac{1}{\epsilon_{0}^{2}} \sqrt{ \frac{\omega_{s} \omega_{i}}{2v_{p}v_{s}v_{i}} } \int_{\chi^{(2)}} \mathrm{d}x \mathrm{d}y \frac{\chi_{jkl}^{(2)}(z)}{n_{j}^{2} n_{k}^{2} n_{l}^{2}} \left[ d_{s}^{j}(x,y) \right]^{\blue{*}} \left[ d_{i}^{k}(x,y) \right]^{*} \left[ d_{p}^{l}(x,y) \right];
    \end{split}
\end{align}
\label{eq:InteractionCoefficients}
\end{subequations}
\end{widetext}
wherein $h(z) \in \{0,1\}$ and $g(z) \in \{-1,0,1\}$ indicate the presence and the sign of the material nonlinearity. The fields $d_{p,s,i}^{k}(x,y)$ ($k \in \{x,y,z\}$) correspond to the cartesian components of the  transverse fields ${\bf d}_{p,s,i}^{\perp}(x,y)$.

Writing the Hamiltonians in eq. (\ref{eq:Hamiltonians}) in terms of the pump amplitude $\mathcal{A}$ and the fields amplitudes $a_{s,i}$ finally allows to calculate the EOM presented in the main text from the Heisenberg equation of motion. Herein, due to the transformation into the different moving frames (introduction of the field operators $\varphi_{j}$ and $a_{j}$), the linear term gives a contribution
\begin{equation}
    \Delta k_{s,i}(\omega) \: = \: \left( \frac{1}{v_{s,i}(\omega)} - \frac{1}{v_{p}(\omega)} \right) (\omega - \omega_{s,i}),
\end{equation}
which describe the walk-off between the pump and the signal/idler fields.

\noindent
\textbf{Calculating interaction parameters-of-interest:}
To calculate parameters-of-interest for the states generated by PDC, we first calculate the second order moment
\begin{equation}
    {\bf M}_{mn} = \bra{\mathrm{vac}} a_{s}(z,\omega_{m}) a_{i}(z,\omega_{n}) \ket{\mathrm{vac}}
\end{equation}
using the transformations of the input creation and annihilation operators in eq. (\ref{eq:DiscretizedEOM}), which can readily be expressed in terms of the ${\bf K}^{j,j}$ ($j \in \{s,i\}$) defined in eq. (\ref{eq:PropagatorEOM}). This allows to write the JSA using the singular value decomposition of ${\bf M} = {\bf V} \left[ \mathrm{\textbf{diag}} \left( \tilde{r}_{1} ,...,\tilde{r}_{N} \right) \right] {\bf W}^{\mathrm{T}}$ as \cite{QuesadaSipe:22}
\begin{equation}
    {\bf J} = {\bf V} \left[ \mathrm{\textbf{diag}} \left( r_{1} ,...,r_{N} \right) \right] {\bf W}^{\mathrm{T}}
\end{equation}
in which the squeezing parameters $r_{n}$ are related to the singular values $\tilde{r}_{n}$ of ${\bf M}$ as $r_{n} = \frac{1}{2} \mathrm{sinh}^{-1} \left( 2 \tilde{r}_{n} \right)$. Assuming vacuum input states, the JSA allows to directly write the states after interaction
\begin{subequations}
\label{eq:StatesVacuumInput}
    \begin{align}
            \ket{\psi_{\mathrm{\tiny PDC}}(z)} = \; &\mathrm{exp} \left[ \sum_{m=1}^{N} \sum_{n=1}^{N} {\bf J}_{mn} a_{s}^{\dagger}(z_{0},\omega_{m}) a_{i}^{\dagger}(z_{0},\omega_{n}) \right. \\
            &\left.- \: \mathrm{H.c.} \right] \ket{\mathrm{vac}}.
    \end{align}
\end{subequations}
From here it is straight-forward to define broadband Schmidt modes for the signal and idler operators from the singular value decomposition of {\bf M}
\begin{subequations}
    \begin{align}
    A_{s,n}^{\dagger} = \sum_{m=1}^{N} {\bf V}_{mn} a_{s}^{\dagger}(z_{0},\omega_{m}) \\
    A_{i,n}^{\dagger} = \sum_{m=1}^{N} {\bf W}_{mn} a_{i}^{\dagger}(z_{0},\omega_{m})
    \end{align}
\end{subequations}
and rewrite the state generated from vacuum as \cite{HeltQuesada:20,QuesadaSipe:22}
\begin{equation}
    \ket{\psi_{\mathrm{\tiny PDC}}(z)} = \mathrm{exp} \left[ \sum_{n=1}^{N} r_{n}  A_{s,n}^{\dagger} A_{i,n}^{\dagger} - \mathrm{H.c.} \right] \ket{\mathrm{vac}}.   
\end{equation}
The singular values $\bar{r}_{n}$ of the moments
\begin{subequations}
    \begin{align}
        \begin{split}
            {\bf N}_{mn}^{s} = \bra{\mathrm{vac}} a_{s}^{\dagger}(z,\omega_{m}) a_{s} (z,\omega_{n}) \ket{\mathrm{vac}}
        \end{split}\\
        \begin{split}
            {\bf N}_{mn}^{i} = \bra{\mathrm{vac}} a_{i}^{\dagger}(z,\omega_{m}) a_{i} (z,\omega_{n}) \ket{\mathrm{vac}}.
        \end{split}
    \end{align}
\end{subequations}
correspond to the expectation values of the photon number down-converted into the respective Schmidt modes. 

For QFC we again consider the transformations between the input and output creation and annihilation operators described by eq. (\ref{eq:DiscretizedEOM}), mediated by the matrices ${\bf K}^{j,j}$ ($j \in \{s,i\}$), and write the joint singular value decompositions \cite{ChristSilberhorn:13}:
\begin{subequations}
    \begin{align}
        \begin{split}
            {\bf K}^{s,i} = {\bf \tilde{V}}^{i,\mathrm{out}} \Bigl[ \mathrm{\textbf{diag}} \left(-\mathrm{sin}(t_{1}) ,...,-\mathrm{sin}(t_{N}) \right) \Bigr] \left[ {\bf \tilde{W}}^{s,\mathrm{in}} \right]^{\mathrm{T}} 
        \end{split}\\
        \begin{split}
            {\bf K}^{i,s} = {\bf \tilde{V}}^{s,\mathrm{out}} \Bigl[ \mathrm{\textbf{diag}} \left(\mathrm{sin}(t_{1}) ,...,\mathrm{sin}(t_{N}) \right) \Bigr] \left[ {\bf \tilde{W}}^{i,\mathrm{in}} \right]^{\mathrm{T}} 
        \end{split}
    \end{align}
\end{subequations}
As before, this allows to express the input and output signal and idler operators using Schmidt modes
\begin{subequations}
    \begin{align}
    A_{s,n}^{\mathrm{in}} &= \sum_{m=1}^{N} {\bf \tilde{W}}^{s,\mathrm{in}}_{mn} a_{s}(z_{0},\omega_{m})\\
    A_{i,n}^{\mathrm{in}} &= \sum_{m=1}^{N} {\bf \tilde{W}}^{i,\mathrm{in}}_{mn} a_{i}(z_{0},\omega_{m})\\
    \left[ A_{s,n}^{\mathrm{out}} \right]^{\dagger} &= \sum_{m=1}^{N} {\bf \tilde{V}}^{s,\mathrm{out}}_{mn} a_{s}^{\dagger}(z_{0},\omega_{m}) \\
    \left[ A_{i,n}^{\mathrm{out}} \right]^{\dagger} &= \sum_{m=1}^{N} {\bf \tilde{V}}^{i,\mathrm{out}}_{mn} a_{i}^{\dagger}(z_{0},\omega_{m})
    \end{align}
\end{subequations}
This allows to calculate the conversion probability of the respective Schmidt modes of the input signal state as $\gamma_{n} = \mathrm{sin}^{2}(t_{n})$.

We note, that, while it is possible to write an expression for the state after interaction with the vacuum using the corresponding nonzero second order moments, this is meaningless, since a QFC type interaction does occur only for non-vacuum input states. 

\noindent
\textbf{Accounting for non-vacuum input states:} To account for non-vacuum inputs we first note that the transformations of the input creation and annihilation operators in eq. (\ref{eq:DiscretizedEOM}) may be written in terms of the time evolution unitary as ${\bf a}_{s,i}(z) = {\bf \mathcal{U}}^{\dagger} {\bf a}_{s,i}(z_{0}) {\bf \mathcal{U}} $. If we further write the input state as $\ket{\psi^{\mathrm{in}}(z_{0})} = f({\bf a}_{s}^{\dagger},{\bf a}_{i}^{\dagger}) \ket{\mathrm{vac}}$, wherein $f$ represents a polynomial or power series of the respective creation operators, this allows to directly write the corresponding state after interaction as \cite{QuesadaSipe:22}
\begin{equation}
\label{eq:NonVacuumInput}
    \begin{aligned}
    \ket{\psi^{\mathrm{in}}(z)} &= {\bf \mathcal{U}} \ket{\psi^{\mathrm{in}}(z_{0})} = {\bf \mathcal{U}} f({\bf a}_{s}^{\dagger},{\bf a}_{i}^{\dagger}) {\bf \mathcal{U}}^{\dagger} {\bf \mathcal{U}} \ket{\mathrm{vac}}\\
    &= f \left( {\bf \mathcal{U}} {\bf a}_{s}^{\dagger} {\bf \mathcal{U}}^{\dagger},{\bf \mathcal{U}} {\bf a}_{i}^{\dagger} {\bf \mathcal{U}}^{\dagger} \right) {\bf \mathcal{U}} \ket{\mathrm{vac}}\\
    &= f\left( \sum_{n=1}^{N} \left[ {\bf \bar{K}}_{mn}^{s,s} \right]^{*} {\bf a}_{s}^{\dagger} + \sum_{n=1}^{N} \left[ {\bf \bar{K}}_{mn}^{s,i} \right]^{*} {\bf a}_{i}^{\orange{\dagger}} \right.,\\
    & \; \; \; \; \; \left. \sum_{n=1}^{N} \left[ {\bf \bar{K}}_{mn}^{i,i} \right]^{*} {\bf a}_{i}^{\dagger} + \sum_{n=1}^{N} \left[ {\bf \bar{K}}_{mn}^{i,s} \right]^{*} {\bf a}_{s}^{\orange{\dagger}} \right) {\bf \mathcal{U}} \ket{\mathrm{vac}};
    \end{aligned}
\end{equation}
wherein the ${\bf \bar{K}}^{j,j}$ ($j \in \{s,i\}$) correspond to the constituents of the (right) inverse ${\bf K}^{-1}$ of the propagator defined in eq. (\ref{eq:DefinitionPropagator}), describing the backward evolved transformations. Appropriate Schmidt modes may be defined in a manner identical to the one above.

\noindent
\textbf{Incorporating losses:} Photon loss of the classical pump field can be included in a straight-forward fashion by taking the total photon number $\mathcal{N}_{p}$ contained in the pump-field amplitude $\Lambda$ (with $\int | \Lambda(z) |^{2} \mathrm{d}z = \mathcal{N}_{p}$) as a position dependent quantity. In this work, we consider only the linear loss, mediated by the loss coefficient $\alpha_{\mathrm{lin}}$ as $\mathcal{N}_{p}(z_{l}) = \mathrm{exp} \bigl[ -\int_{z_{0}}^{z_{l}} \alpha_{\mathrm{lin}}(z) \mathrm{d}z \bigr] \mathcal{N}_{p}(z_{0})$, but note, that nonlinear losses may be included by adding the respective coefficients.

To incorporate losses into the signal and idler modes we insert fictional beamsplitters at the locations $z_{l}$ of the Trotter-Suzuki expansion, each coupling waveguide modes to the environment, which is prepared in the vacuum and described by operators ${\bf c}$,${\bf c}^{\dagger}$. The corresponding interaction can be written as 
\begin{equation}
\label{eq:BSLoss}
    {\bf a}_{s,i}(z_{l} + \Delta z_{l}) = {\boldsymbol \eta}_{s,i} \, {\bf a}_{s,i}(z_{l}) + \sqrt{\mathbb{1} - {\boldsymbol \eta}_{s,i} \: {\boldsymbol \eta}_{s,i}^{\dagger}} \:{\bf c},
\end{equation}
wherein ${\boldsymbol \eta}$ is a frequency dependent loss matrix. From here on, it can be shown, that the different second order moments transform as \cite{HeltQuesada:20,QuesadaSipe:22}
\begin{subequations}
    \begin{align}
    {\bf M}(z_{l} + \Delta z_{l}) &= {\boldsymbol \eta}_{s} {\bf M}(z_{l}) {\boldsymbol \eta}_{i}^{\mathrm{T}}\\
    {\bf N}^{s,i}(z_{l} + \Delta z_{l}) &= {\boldsymbol \eta}_{s,i}^{*} {\bf N}^{s,i}(z_{l}) {\boldsymbol \eta}_{s,i}^{\mathrm{T}},
    \end{align}
\end{subequations}
wherein, for simple, linear losses, ${\boldsymbol \eta}$ can be written as a diagonal matrix ${\boldsymbol \eta}_{mm}= \mathrm{exp} \bigl[ -\frac{1}{2} \int_{z_{0}}^{z_{l}} \alpha_{\mathrm{lin}}(z,\omega_{m}) \mathrm{d}z \bigr]$.

To obtain the corresponding transformation for the input operators, we insert the beamsplitter transformations (eq. \ref{eq:BSLoss}) in between the respective factors of the Trotter-Suzuki expansion in eq. (\ref{eq:DefinitionPropagator}). If the losses are assumed uniform across both the signal and idler frequency-bands, and the environment modes of subsequent interactions are uncorrelated and prepared in the vacuum, the corresponding transformation (described by eq. (\ref{eq:PropagatorEOM})) can be written as
\begin{equation}
\label{eq:Losses}
    \begin{aligned}
        \begin{pmatrix}
        {\bf a}_{s}(z) \\
        {\bf a}_{i}^{\blue{\dagger}}(z)
        \end{pmatrix}
        \; = \; &\eta_{\mathrm{tot}} {\bf K}(z,z_{0})
        \begin{pmatrix}
        {\bf a}_{s}(z_{0}) \\
        {\bf a}_{i}^{\blue{\dagger}}(z_{0})
        \end{pmatrix}\\
        &+ \sqrt{1- \eta_{\mathrm{tot}}^{2} }
        \begin{pmatrix}
        {\bf c}_{s,\mathrm{eff}}(z) \\
        {\bf c}_{i,\mathrm{eff}}^{\blue{\dagger}}(z)
        \end{pmatrix}
    \end{aligned}
\end{equation}
wherein $\eta_{\mathrm{tot}}$ corresponds to the loss over the entire propagation length $\eta_{\mathrm{tot}} = \prod_{l} \eta_{l} = \mathrm{exp} \bigl[ -\frac{1}{2} \int_{z_{0}}^{z} \alpha_{\mathrm{lin}}(z) \mathrm{d}z \bigr]$; ${\bf c}_{j,\mathrm{eff}}$ are effective environment annihilation operators defined as the normalized sum of the contributions describing the individual interactions. To calculate the relations describing the transformation of the non-vacuum input states in eq. (\ref{eq:NonVacuumInput}), is useful to rewrite eq. (\ref{eq:Losses}) using the matrix ${\bf K}^{\mathrm{loss}} = \left[ \eta_{\mathrm{tot}} {\bf K} \: \big| \: \sqrt{1- \eta_{\mathrm{tot}}^{2} } {\bf \mathbb{1}} \right]$. This description allows calculation of the right inverse $\left[ {\bf K}^{\mathrm{loss}} \right]^{-1} = \left[ {\bf K}_{\mathrm{system}}^{\mathrm{loss}} \: \big| \: {\bf K}_{\mathrm{env}}^{\mathrm{loss}} \right] $, describing the backward evolution required for eq. (\ref{eq:NonVacuumInput}). 

\bibliography{References}

\clearpage
\setcounter{figure}{0}
\makeatletter 
\renewcommand{\thefigure}{S\@arabic\c@Fig}
\makeatother

\setcounter{equation}{0}
\makeatletter 
\renewcommand{\theequation}{S\@arabic\c@Eq}
\makeatother

\setcounter{table}{0}
\makeatletter 
\renewcommand{\thetable}{S\@arabic\c@table}
\makeatother

\end{document}